\documentclass{aa}  

\usepackage{graphicx}
\usepackage{txfonts}
\usepackage{natbib}
\defcitealias{FA21}{FA21}

\begin{document}

   \title{On the metal-poor edge of the Milky Way "thin disc".}


   \author{Emma Fern\'andez-Alvar
          \inst{1,2,3},
          Georges Kordopatis\inst{3},
          Vanessa Hill\inst{3},
          Giuseppina Battaglia\inst{1,2},
          Carme Gallart\inst{1,2},
          Isaure Gonz\'alez Rivera de la Vernhe\inst{3},
          Guillaume Thomas\inst{1,2},
          Federico Sestito\inst{4},
          Anke Ardern-Arentsen\inst{5},
          Nicolas Martin\inst{6,7},
          Akshara Viswanathan\inst{8},
          \and
          Else Starkenburg\inst{8}
          }

   \institute{Departamento de Astrof\'isica, Universidad de La Laguna, E-38206 La Laguna, Tenerife, Spain
      			  \and
				Instituto de Astrof\'isica de Canarias, E-38200 La Laguna, Tenerife, Spain
   			     			  \and
   			  Universit\'e C\^ote d'Azur, Observatoire de la C\^ote d'Azur, CNRS, 06300, Laboratoire Lagrange
   			     			  \and
   			  Department of Physics \& Astronomy, University of Victoria, Victoria, BC, V8P 1A1, Canada
   			     			  \and
   			  Institute of Astronomy, University of Cambridge, Madingley Road, Cambridge CB3 0HA, UK
   			     			  \and
   			  Universit\'e de Strasbourg, CNRS, Observatoire astronomique de Strasbourg, UMR 7550, F-67000 Strasbourg, France
   			     			  \and
   			     			  Max-Planck-Institut f\"{u}r Astronomie, K\"{o}nigstuhl 17, D-69117 Heidelberg, Germany
   			     			  \and
   			  Kapteyn Astronomical Institute, University of Groningen, Landleven 12, 9747 AD Groningen, The Netherlands\\
              \email{emmafalvar@gmail.com}
             }

   \authorrunning{Emma Fern\'andez-Alvar et al.}
   \date{}

 
  \abstract
   {The emergence of the disc in our Galaxy and the relation of the thick and thin disc formation and evolution is still a matter of debate. The chemo-dynamical characterization of disc stars is key to resolve this question, in particular at parameter regimes where both disc components overlap, such as the region around [Fe/H] $\sim$ $-0.7$ corresponding to the thin disc metal-poor end.}
   {In this paper we re-assess the recent detection of a metal-poor extension of stars moving with thin-disc-like rotational velocities between -2 < [Fe/H] < -0.7 that was made based on metallicity estimates obtained from photometric data and their rotational velocity distribution. }
   {We explore the chemo-dynamical properties of metal-poor stars within the recent Gaia third data release (DR3), which includes the first catalogue of metallicity estimates from the Radial Velocity Spectrometer (RVS) experiment. We complement them with the two largest high-resolution ($\lambda/d\lambda$ > 20,000) spectroscopic surveys available, the GALAH DR3 and the APOGEE DR17.}
   {We confirm that there are high angular-momentum stars moving in thin-disc-like orbits, i.e., with high angular momentum $\rm L_{z}/J_{tot}$ > 0.95, and close to the Galactic plane, $\rm |Z_{max}|$ < 750 pc, reaching metallicity values down to [Fe/H] $\sim-1.5$. We also find tentative evidence of stars moving on such orbits at lower metallicities, down to [Fe/H] $\sim -2.5$, although in smaller numbers. Based on their chemical trends the fast rotators with [Fe/H] < -1 would have formed in a medium less chemically evolved than the bulk of the thick disc. Fast rotators with chemical abundances typical of the thin disc appear at metallicities between -1 < [Fe/H] < -0.7.}
   {}

   \maketitle


\section{Introduction}

In the past decades, analysis of the Milky Way (MW) have clearly established that most stars in our Galaxy move in a disc configuration following a double density law, leading to a separation in distinct thin and thick disc components \citep{gilmore83}. Further chemo-dynamical analysis revealed that both components also differentiate themselves chemically (e.g., in the [$\alpha$/Fe] vs. [Fe/H] abundance space) and dynamically, distributing in different proportions over the Galactic plane and from the Galactic center \citep{juric2008, kordopatis2015,hayden15,queiroz20}. 

Thin disc stars, i.e., those moving closer to the plane with low eccentricities, display an overall low [$\alpha$/Fe] enhancement, and a metallicity distribution centered around the solar value \citep{rb14, hayden14} which ranges from supersolar values down to [Fe/H] $\sim$ $-0.7$ on its metal-poor end \citep{bensby14, fuhrmann17}. On the contrary, the thick disc reaches larger distances from the plane, rotating with a lower angular momentum. Its stars show a higher [$\alpha$/Fe] enhancement and a poorer metallicity content than the thin disc, with a broader metallicity distribution centered around [Fe/H] $\sim$ -0.55 \citep{gilmore89, katz11, kordopatis11b} extending towards much lower values -- several works have even reported the identification of a very metal-poor population moving in orbits compatible with the thick disc, the so-called metal-weak thick disc, down to metallicities [Fe/H] $\sim$ -2.5 \citep{norris85b, beers02, ruchti11, kordopatis13b, carollo19}, although its link with the more metal-rich thick disc is still a matter of debate \citep{mardini22}. Even if in general the chemo-dynamical properties of the thin and the thick discs are different, there is a large overlap between them and the thin/thick disc classification based on their chemistry does not correspond exactly to the classification based on their dynamics. This fact makes it difficult to understand the disc components origin.

Cosmological simulations of Milky-Way-like galaxies point to several formation scenarios that could explain the double disc characteristics: the formation of a dynamically hot disc after an early stage of turbulent accretion of stellar substructures into the Milky Way halo (e.g., \citealt{jw83}; \citealt{brook04}; \citealt{bird13}); or the formation of the thick disc from dynamical heating of a thinner component caused by mergers of satellite galaxies into the Galaxy (\citealt{quinn93}; \citealt{font01}; \citealt{vh08}, among others); or by depositing stars \citep{abadi03}. To this day, it is not clear when the thin disc started forming. In any case, the first stages of Galaxy formation seem to have been dominated by a high merger activity \citep{elbadry18} which would prevent the formation of a disc that would be stable over time. Thus, finding old stars (and, consequently, with a very low metallicity content) moving in cold disky orbits is unexpected and puzzling. 

Recent studies find more and more evidences of very ([Fe/H] < -2), extremely ([Fe/H] < -3) and even ultra metal-poor stars ([Fe/H] < -4) with a large angular momentum and confined relatively close to the Galactic plane \citep{sestito19, sestito20, sestito21, dimatteo20, venn20, cordoni21, dovgal23}, most of them moving in thick-disc-like orbits. \citet{sestito21} and \citet{santistevan21} investigated their possible origin with cosmological simulations concluding that some of the stars from accreted structures that merged early and became the Milky Way building blocks and/or later mergers in particular directions would end up on relatively planar orbits. Another kind of explanation was recently proposed by \citet{dillamore23}. Exploring particle simulations they showed that a rotating bar can give the stellar halo a net spin leading to an asymmetric prograde angular momentum distribution with some stars moving close to the plane on disc-like orbits. \citet{yuan23} and \citet{li23} also presented how due to a rapidly decelerating bar some bulge stars gain rotation because they are trapped in co-rotating regions and move outwards on prograde planar orbits. However, they concluded that the fraction of stars affected by this mechanism is too small to account for all the metal-poor rotators detected. Thus, their origin is still unclear.

A recent analysis by \citet{FA21} (here after FA21) of stars observed as part of the Pristine survey \citep{starkenburg17, martin23} towards the Galactic anticenter revealed a stellar population moving with rotational velocities not just disc-like but typical of the thin disc at metallicities down to [Fe/H] $\sim$ -2. The metallicities used in the \citetalias{FA21} analysis were derived photometrically with the Pristine filter centered on the CaHK doublet at $\sim$ 3900 $\AA$, which is very sensitive to a variation of the global metallicity of the star. The aim of the present work is to verify with more accurate spectroscopic metallicities, and a more complete dynamical characterization, the existence of high angular-momentum stars at very low metallicities, i.e., -2 < [Fe/H] < -0.7, and investigate their possible linked origin with the chemically defined thin disc, i.e., the low-[$\alpha$/Fe] population. 

The Gaia third data release (DR3; \citealt{gaia16b, gaia23j}) provides the first determinations of stellar parameters and chemical abundances obtained from the Radial Velocity Spectrometer (RVS; \citealt{cropper18, rb22a}). This is the largest existing sample of stellar parameters and individual abundances homogeneously derived for Milky Way stars from medium-resolution spectra ($\lambda/d\lambda$ $\sim$ 11,500). It comprises a database of 5.6 million stars, including several $10^{4}$ stars with [Fe/H] < -1. Combined with the highly accurate astrometric measurements, this constitutes a extremely valuable sample to explore the metal-poor tail of the thin disc. In addition, we also investigate the two largest high-resolution spectroscopic surveys available, the GALAH DR3 ($\lambda/d\lambda$ $\sim$ 28,000; \citealt{buder21}) and APOGEE DR17 ($\lambda/d\lambda$ $\sim$ 22,500; \citealt{sdssdr17}). 

This paper is structured as follows. We present the data and the quality selection criteria applied in Section 2. We explore the metallicity distribution function (MDF), chemical abundance trends and orbital parameters of dynamically selected thin-disc-like stars in Section 3. Finally, we summarise and discuss our results in Section 4.


\section{Data selection.}

\begin{figure}
\includegraphics[scale=0.6, trim = 250 400 0 90]{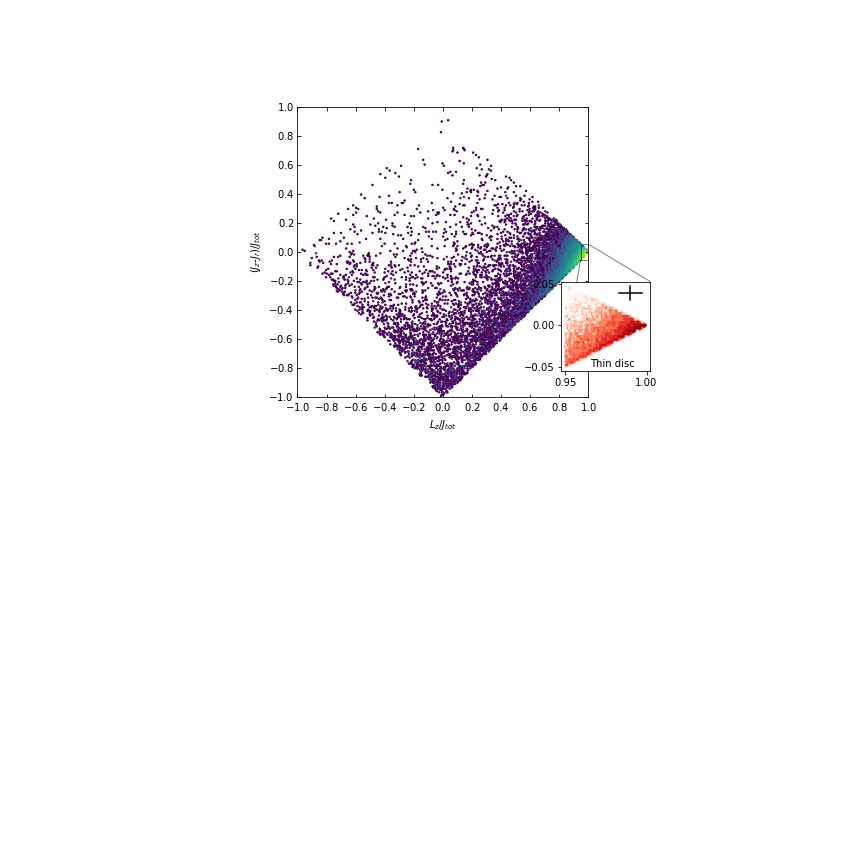}
\caption{Action space (normalized to the sum of the three actions of motion, \textit{Jtot}) of Gaia RVS stars with calibrated [Fe/H] $<$ -0.55, with a zoom-in of the area that we use to select our thin disc sample. The xaxis displays the normalized angular momentum, which is equivalent to the azimuthal action of motion and the yaxis the normalized difference of vertical action and radial action. Their mean uncertainties are displayed in black on the top right corner. }
\label{actions}
\end{figure}


We select stars within the Gaia DR3 RVS, GALAH DR3 and APOGEE DR17 spectroscopic databases with estimated metallicities lower than [Fe/H]\footnote{Both Gaia RVS and APOGEE provide two sets of iron abundances: [M/H], calculated using all the spectral range available, and [Fe/H], estimated by taking into account only individual iron lines. In this work we consider the [M/H] parameter. In the case of GALAH they only provide a unique set of [Fe/H] values. For convenience, we will refer to them all as [Fe/H] from now on.} < -0.55, where the probability of finding stars in thin-disc orbits is very low based on the standard thin disc MDF (e.g., \citealt{bensby14}), even considering the disc metallicity gradients with both radial and vertical distance ($\sim -0.06\pm\ 0.001$ dex $\rm kpc^{-1}$  and $\sim -0.26\pm\ 0.01$ dex $\rm kpc^{-1}$, respectively -- e.g.\citealt{kordopatis20, imig23}).  

We search for the best quality data among the Gaia DR3 RVS database by following the selection suggested by \citet{rb22a} for metal-poor stars in their section 10.5. Apart from the criteria they provided, we conservatively include additional constraints to reject hot and cool stars for which the metallicity determination could be less reliable because of their proximity with the limits of the \textit{GSP-Spec} model grid used in their analysis. Thus, we remove stars reported in the \textit{Astrophysical-Parameters} table as $spectraltype\_esphs$ type O, B and A; we restrict our stars to stellar parameters between  4250 < $T_{\rm eff}$ [k] < 6000 and 0 < $\log g$ [dex] < 4.75; and we consider only stars with the $flag\_extrapol$ < 2. We provide a table of the flags applied in Appendix \ref{flags}. We calibrate the RVS metallicities with the $\log g$ dependent third order polynomial suggested by \citet{rb22a} and keep stars with calibrated metallicities lower than [Fe/H] < -0.55. 

Because of the small wavelength range covered by the RVS spectra and the fact that it is dominated by the Ca II triplet (at 8498 \AA, 8542 \AA\ and 8662 \AA), some degeneracies between the [Fe/H] and [$\alpha$/Fe] determinations could lead to systematic errors when deriving these chemical abundances (e.g. \citealt{kordopatis11a, matsuno22}). For instance, an overestimation of [$\alpha$/Fe] would translate in underestimating [Fe/H]. In that case, some metal-rich thin disc stars could contaminate our sample of interest, below [Fe/H] < $-1$. These would show large [$\alpha$/Fe] enhancements. We evaluate this issue by comparing the Gaia RVS [Fe/H] of our selected sample with the [Fe/H] of stars in common within GALAH and APOGEE databases. The comparison showed that we avoid underestimated [Fe/H] by removing stars at [Fe/H] > -2 with $flags\_gspspec$ chain character number 7 equal to 2. The details of this comparison are shown in Appendix \ref{ab_comp}.

Among the GALAH data, we filter those with unreliable stellar parameters ($flag\_sp$ = 0) and with unreliable metallicities ($flag\_fe\_h$ = 0) as suggested by \citet{buder21}. During our analysis, we detected a decreasing trend of the metallicity with $T_{\rm eff}$ in stars with $T_{\rm eff}$ > 6500 K. Overestimating $T_{\rm eff}$, for example because of a sub-optimal broadening model, would lead to underestimate significantly [Fe/H]. To avoid this problem, we focus on cooler stars only and reject those with $T_{\rm eff}$ differences between photometric and spectroscopic determinations (both provided by the GALAH collaboration -- \citealt{buder21})  larger than 300 K. We add an additional cut removing stars with a large velocity broadening, $v_{\rm broad}$ > 15 $\rm km\ s^{-1}$, that could be related to unresolved binaries or fast spinning stars, and would also lead to inaccurate [Fe/H] measurements\footnote{For Gaia RVS metallicities this issue was taking care of through the $v_{\rm broad}$ flags as shown in Appendix \ref{flags}.}.

In the case of APOGEE data, we select only stars with ASPCAPFLAG bits 14 to 42 equal to 0, STARFLAG bits 1 to 26 equal to 0, and EXTRATARG bits 2 and 4 equal to 0. We also exclude those objects with the $apogee1\_target1$, $apogee1\_target2$, $apogee2\_target1$ and $apogee2\_target2$ classifying them as clusters, streams, dwarf galaxies, sky, telluric, binaries, radial velocity variables, the bar, extended objects and the TriAnd, GASS and A13 disc structures. In addition, we select only stars with 3500 < $T_{\rm eff}$ [K] < 6500, and 1 < $\log g$ [dex] < 3.5 where the determination of stellar parameters, including metallicity, are more reliable \citep{jonson2020}. For consistency with the Gaia RVS final metallicity upper limit we restrict our GALAH and APOGEE samples to stars with [Fe/H] < -0.55.

In order to select stars in thin-disc-like orbits (see Section \ref{results}) we consider the actions of motion of the stars, $\rm J_{r}$, $\rm J_{\phi}$ and $\rm J_{z}$\footnote{$\rm J_{\phi}$ is identical to the angular momentum, $\rm L_{z}$. We will refer to it as $\rm L_{z}$ from now on.}, which are invariants with time (for a static potential in a galaxy in equilibrium) and can offer a better classification of the stellar motion than present day velocities. We also evaluate their rotational velocity, $V_{\phi}$, maximum distance from the plane reached in their orbits, $\rm |Z_{max}|$, and guiding radius, $\rm R_{guide}$\footnote{$\rm R_{guide} = (R_{apo}+R_{peri})/2$, being $\rm R_{apo}$ and $\rm R_{peri}$ the apocenter and pericenter of a stellar orbit, respectively.}. We derived the velocities and actions of motion as in \citet{kordo23} (see their section 3.1) considering the spectroscopic GaiaRVS/GALAH/APOGEE radial velocities, Gaia DR3 proper motions and distances calculated by \citet{bj21}.

We impose some additional quality data cuts to minimize parameter uncertainties that can result into non-physical properties for some disc stars. To remove possible binaries we discard the stars for which the measured GALAH/APOGEE radial velocity is not in agreement with the Gaia measurement, within the uncertainties. We also exclude likely binaries based on the Gaia ruwe parameter (ruwe > 1.4 as suggested by \citealt{lindegren21}). We restrict ourselves to precise distances by removing stars with parallax relative uncertainties greater than 20\%. Finally, we do not take into account those for which our final total velocity exceeds the escape velocity of the Galaxy, 600 $\rm km\ s^{-1}$, calculated using the \citet{McMillan17} potential as in \citet{kordo23}. These selection criteria result in samples of 41682 Gaia DR3 RVS stars, 21544 GALAH DR3 stars, and 16404 APOGEE DR17 stars.

\section{Metal-poor stars on thin-disc orbits.}
\label{results}

\begin{figure}
\includegraphics[scale=0.6, trim = -5 50 -15 80]{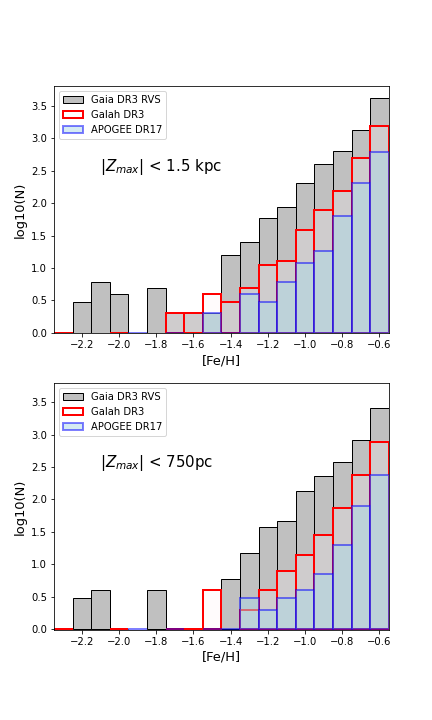}
\caption{Metallicity distribution functions, in logarithmic scale, derived from our thin disc samples of Gaia DR3 RVS, GALAH and APOGEE stars.}
\label{mdf_3}
\end{figure}

\begin{figure*}
\includegraphics[scale=0.5, trim = 20 100 0 100]{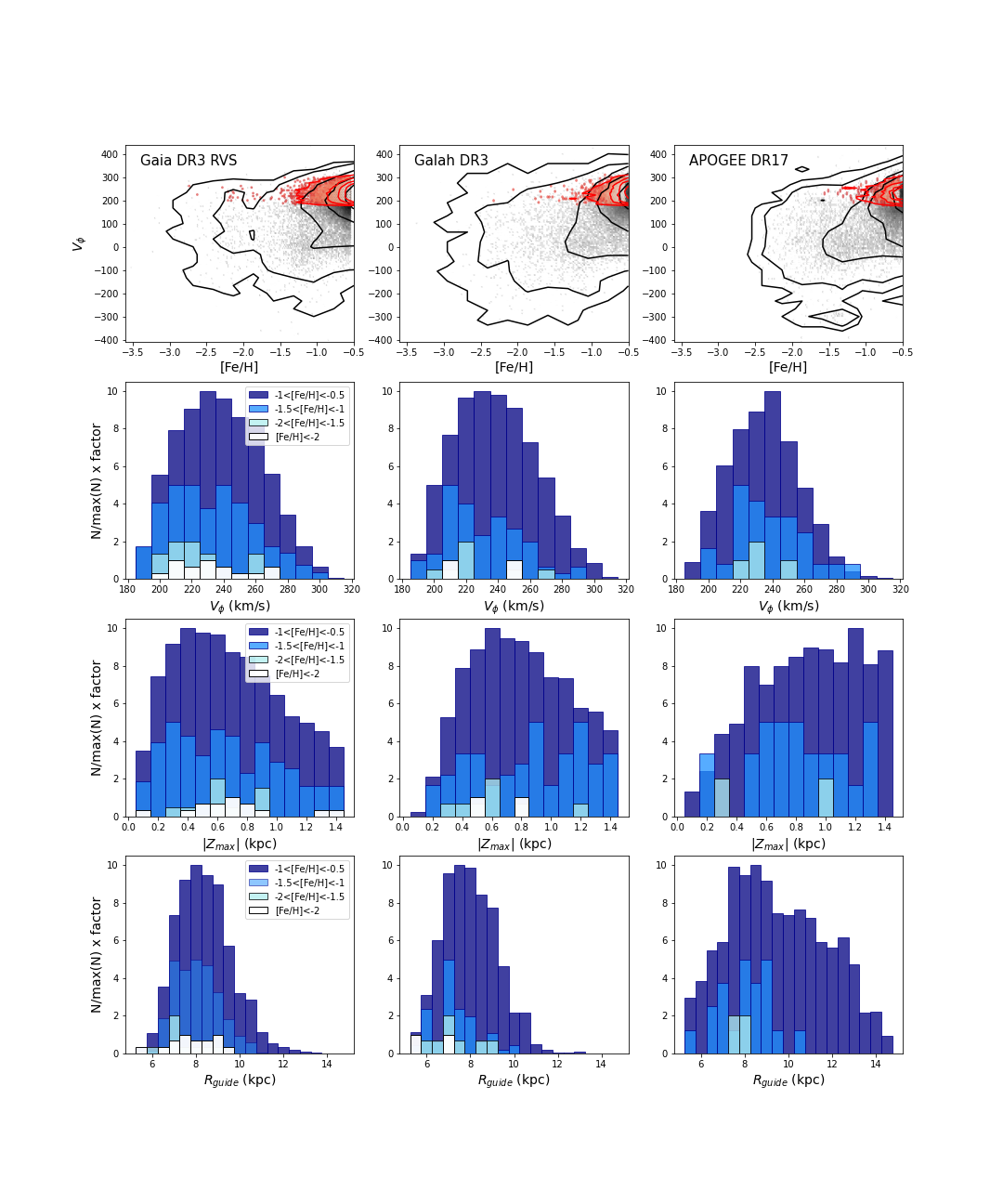}
\caption{First row: rotational velocity, $\rm V_{\phi}$, as a function of the metallicity, [Fe/H], color coded by density scale, for the thin disc selections. Thin disc stars are shown in red scale overplotted over the rest of the sample in grey scale. Contour plots of the 50\%, 75\%, 90\%, 99\%, and 99.9\% (the latter only for the metal-poor thin disc) are overplotted following the same color code. Second row: $\rm V_{\phi}$ distributions of thin disc stars at four different ranges of metallicity (from dark blue to white, [Fe/H] < -0.5, [Fe/H] < -1, [Fe/H] < -1.5, [Fe/H] < -2), normalized to the maximum number of stars and multiplied by a factor to help the visualization (from dark blue to white, x10,x5,x2 and x1). Third and fourth rows:  Same as middle panels but for $\rm |Z_{max}|$ and $\rm R_{guide}$ distributions, respectively. From left to right we display the Gaia DR3 RVS, GALAH and APOGEE stellar samples. }
\label{vphi_feh}
\end{figure*}

Figure \ref{actions} represents the combined vertical and radial actions $\rm (J_{z}-J_{r})$ as a function of the azimuthal action $\rm L_{z}$, normalized to the sum of the three, $\rm J_{tot} = J_{z} + |L_{z}| + J_{r}$. The stars of interest, i.e., moving on circular orbits on the Galactic plane, would be those with $\rm L_{z}/J_{tot} \sim 1$ and $\rm (J_{z}-J_{r})/J_{tot} \sim 0$. We select stars on thin-disc-like orbits by choosing those with $\rm L_{z}/J_{tot}$ > 0.95, which have a $\rm |(J_{z}-J_{r})/J_{tot}|$ < 0.05. This selection still includes some stars reaching large distances from the plane (up to $\rm |Z_{max}|$ $\sim$ 3 kpc -- see left panel of Figure \ref{apc_fig}). To restrict our sample closer to the plane, we keep stars with $\rm |Z_{max}|$ < 1.5 kpc. Whereas this limit is still high for stars to be considered thin disc, we nevertheless adopt it, in order to be consistent with the location where most of the Pristine stars in \citetalias{FA21} were observed. Based on Figure \ref{apc_fig}, most of them are confined within $\rm |Z_{max}|$ < 750 pc, which is consistent with the limits of the geometrical thin disc (with a scale height of 300 pc measured in the solar neighbourhood -- \citealt{bd16}).  The analysis below confirms that the bulk of these stars indeed display $\rm |Z_{max}|$ significantly lower than this 1.5kpc limit, irrespective of their metallicity.  

We obtain 12889 Gaia DR3 RVS stars, 4097 GALAH DR3 stars, 1840 APOGEE DR17 stars that fulfill our orbital and chemical criteria. Among those, there are 211 stars in common between Gaia RVS and APOGEE, 211 between Gaia RVS and GALAH, and only 9 in common between the three surveys. As expected based on the selection criteria, the eccentricities, $ecc$, of the resulting samples are very low, distributing uniformingly between 0 < $ecc$ < 0.3 and showing no trends with metallicity (see right panel of Figure \ref{apc_fig}). We evaluate the spatial distribution covered by our thin disc samples in Figure \ref{location} in Appendix \ref{app_loc}.

We derive the metallicity distribution function (MDF) of our selected fast rotators, which are displayed in Figure \ref{mdf_3} on a logarithmic scale\footnote{The MDF derived with APOGEE data is already cleaned from stars with a likely accreted origin, as suggested by their chemical trends, see Section \ref{chem_trends}}. The top panel corresponds to the MDF for our final selection with $\rm |Z_{max}|$ < 1.5 kpc, and the bottom panel shows the MDF of those stars that reach only up to $\rm |Z_{max}|$ < 750 pc, which resembles the previous one, although the lack of stars below [Fe/H] < $-1.5$ dex is more evident for the latter. We asses the impact of the uncertainties of the orbital parameters used to make the selection in our results. Figure \ref{mdf_ap} in Appendix \ref{mdf_ap_sec} shows the same shape as in Fig.~\ref{mdf_3}, albeit a stricter selection, excluding all the stars for which the orbital parameters leave them out of the selection described above (considering their higher limit uncertainty).

The three surveys show similar MDFs: a continuous decrease down to [Fe/H] $\sim$ -1.5 and, although in much lower numbers, the existence of fast rotators down to [Fe/H] $\sim$ -2.5. In order to compare quantitatively the MDFs of the three surveys, the percentage of stars below -1, -1.5 and -2 with respect to those below -0.55 (and $\rm |Z_{max}|$ < 1.5 kpc) are shown in Table \ref{table1}. The coherence in the MDFs derived from surveys that used different methodologies to determine stellar metallicities confirms the existence of this fast rotating metal-poor tail. In addition, the detection of this population does not depend on the specific selection functions of these three surveys nor on a particular location in the Galaxy, as it is shown in Figure \ref{location} of Appendix \ref{app_loc}. In particular, they are not restricted to the anticenter region where \citetalias{FA21} had detected initially.

\begin{table}
\caption{Number of stars below [Fe/H] < -0.55, -1, -1.5 and -2 and the corresponding percentage (in brackets) with respect to the number of stars with [Fe/H] < -0.55.}             
\label{table1}      
\centering                          
\begin{tabular}{l c c c c}        
\hline\hline                 
Survey & $< -0.55$ & $< -1$ & $< -1.5$ & $< -2$ \\    
\hline                        
 Gaia RVS &  12889 & 415 (3.22) & 27 (0.21) & 15 (0.12) \\      
 GALAH & 4097 & 80 (1.95) & 10 (0.24) & 2 (0.05) \\
 APOGEE & 1788 & 26 (1.45) & 2 (0.11) & 0 (0) \\
\hline                                   
\end{tabular}
\end{table}

Top panels in Figure \ref{vphi_feh} show the stellar orbital rotational velocity as a function of metallicity. This is the parameter space in which \citetalias{FA21} identified the potential metal-poor thin disc tail. The three top panels show, from left to right, all stars with [Fe/H] < -0.55 in Gaia DR3 RVS, GALAH DR3 and APOGEE DR17 (after the quality selection). The metal-poor fast rotators are overplotted in a red density scale, on top of the gray density scale that shows the complete metal-poor sample. Contour plots of the 50\%, 75\%, 90\%, 99\%, and 99.9\% (the latter only for our fast rotators) are overplotted following the same color code.

The distributions of our three selected samples resemble qualitatively the one displayed in \citetalias{FA21}: a large fraction between -0.55 and -1.5, and a significant drop at lower metallicities. However, contrary to what was observed in \citetalias{FA21}, the number of stars below -1.5 is much lower. This fact could be due to a large fraction of underestimated [Fe/H] within the Pristine photometric metallicity catalogue that would be artificially increasing the number of stars down to [Fe/H] $\sim$ -2 (see Gonz\'alez-Rivera et al. in prep). Another explanation could be that the stars with high $\rm V_{\phi}$ in Pristine do not enter the strict thin disc selection based on actions applied in the present work. Either way, the number of stars detected in these surveys that move in thin-disc-like orbits at metallicities below [Fe/H] < -1.5 is very low compared with the rest of the stellar sample at the same metallicity.

The following rows in Figure \ref{vphi_feh} show the distributions of $\rm V_{\phi}$, $\rm |Z_{max}|$ and $\rm R_{guide}$, from top to bottom respectively, in four different metallicity ranges ($-1$ < [Fe/H] < $-0.5$, $-1.5$ < [Fe/H] < $-1$, $-2$ < [Fe/H] < $-1.5$ and [Fe/H] < $-2$). We see that the $\rm V_{\phi}$ distributions do not strongly depend on the metallicity range considered and they are compatible with a thin disc velocity distribution ($<V_{\phi}> \sim 240$ km $\rm s^{-1}$ -- \citealt{gravity19}), although the stellar velocity distribution at -1 < [Fe/H] < -0.55 peaks at slightly higher velocities than stars at lower metallicities. 

There is a large fraction of stars with -1 < [Fe/H] < -0.55 confined closer than 500 pc from the plane, and at lower metallicities the peak of the distribution moves slightly towards larger $\rm |Z_{max}|$, around 500-750pc, although some stars are also very close to the plane, with $\rm |Z_{max}|$ $\sim$ 100 pc. Even if the $\rm |Z_{max}|$ upper limit of 1.5 kpc imposed in our selection criteria is a bit large to be considered a geometric thin disc, looking at the resulting $\rm |Z_{max}|$ distributions we confirm that most of our stars are confined in thin-disc-like orbits, very close to the plane.

Because of our criteria selecting stars in almost circular orbits, the $\rm R_{guide}$ must be similar to their current galactocentric radius and, thus, their distribution depend on the selection function of each survey. APOGEE covers larger distances than the other surveys towards the outer disc. Interestingly, it shows that metal-rich stars $\rm R_{guide}$ reach up to 14kpc but stars at [Fe/H] < $-1$ are confined within radius below $\sim$ 10 kpc, as if they would not belong to the outer disc.

\subsection{Chemical trends to shed light on their origin.}
\label{chem_trends}

\begin{figure*}
\includegraphics[scale=0.5]{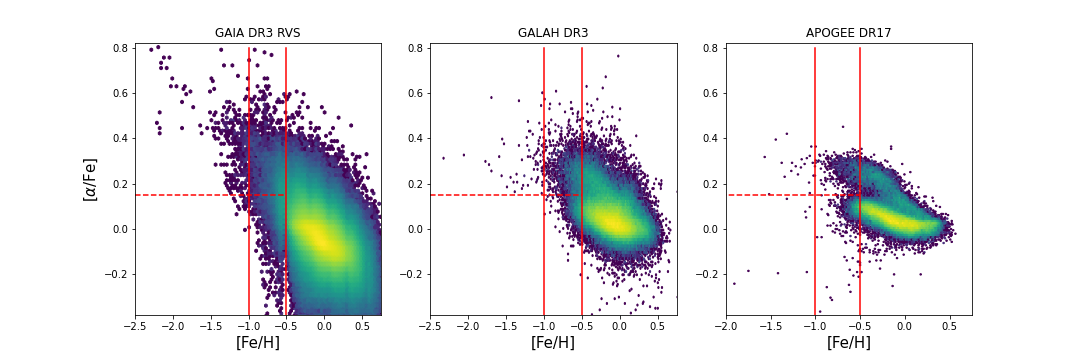}
\caption{[$\alpha$/Fe] ratios as a function of [Fe/H] estimates derived from Gaia RVS (left panel), GALAH (middle panel) and APOGEE (right panel) for stars at all metallicities moving in thin-disc-like orbits based on our selection criteria. Two red lines at [Fe/H] = -0.5 and [Fe/H] = -1 help to visualize the location of our metal-poor thin disc stars in this chemical space. The red dashed line at 0.15 indicates the separation in high- and low-$\alpha$.}
\label{afe}
\end{figure*}

\begin{figure*}
\includegraphics[scale=0.5]{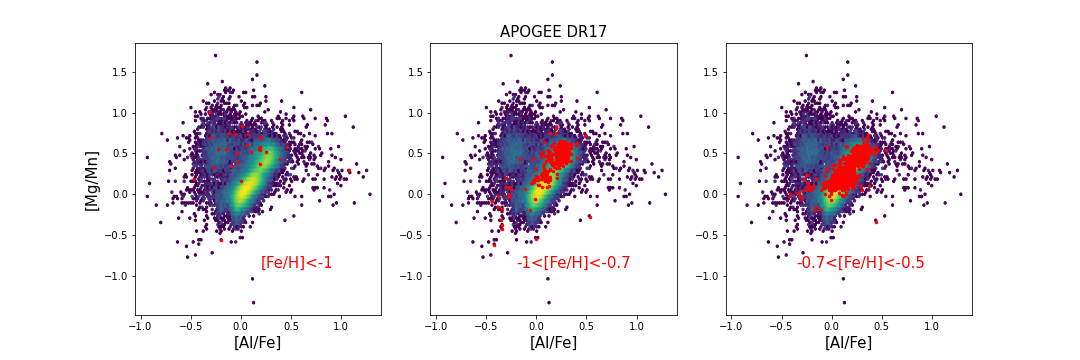}
\caption{[Mg/Mn] vs. [Al/Fe] chemical ratios of APOGEE stars verifying our selection criteria, color coded on stellar density. Overplotted in red are fast rotators with metallicities below [Fe/H] < -1 (left panel), -1 < [Fe/H] < -0.7 (middle panel), and -0.7 < [Fe/H] < -0.55 (right panel).}
\label{mgmn_alfe}
\end{figure*}

As mentioned before, previous studies have shown that at metallicities [Fe/H] > $-0.7$, Milky Way stars moving in thick-disc-like orbits tend to have high-[$\alpha$/Fe] ratios (chemical thick disc), and stars moving in thin-disc-like orbits show low-[$\alpha$/Fe] ratios (chemical thin disc). Figure \ref{afe} displays the [$\alpha$/Fe] vs. [Fe/H] space for stars in the three surveys following our thin-disc selection criteria, at all [Fe/H]. APOGEE abundances clearly show that our high angular-momentum selection is not only comprised by the low-$\alpha$ population, but there is also a fraction of high-$\alpha$ stars;  there are also some stars with a very low [$\alpha$/Fe] content, [$\alpha$/Fe] $\sim$ -0.2, which may point to an accreted origin \citep{tolstoy09, das20}. Both GALAH and Gaia RVS show consistent trends with APOGEE, although more scattered and not split into two sequences\footnote{This is likely due to the fact that GSPspec's alpha is mostly a [Ca/Fe] measurement, which does not separate high-alpha and low-alpha populations clearly, see \citep{prantzos23, mikolaitis14}}. for the latter. 

We roughly separate the stars into  high-$\alpha$ and low-$\alpha$, following the observed split in the APOGEE abundances at [$\alpha$/Fe]$\sim$0.15.
The three catalogues show that among stars with -1 < [Fe/H] < -0.55, around 50\% are high- and low-$\alpha$, but below [Fe/H] < -1, most of our stars tend to be high-$\alpha$.

We also explore the [Mg/Mn] vs. [Al/Fe] which is another chemical space where the Galactic components also differentiate chemically \citep{hawkins15, horta21}. In their Figure 2, \citet{horta21} illustrated the chemical evolution followed by a Milky-Way-like system and a GES-like galaxy on this space. The upper left region is populated by the first stars formed in both systems, when the chemical enrichment is still early and unevolved. Later, due to a fast but intense star formation rate, the Milky-Way follows an increase of [Al/Fe] due to the contribution of AGBs while the [Mg/Mn] remains high (thick disc) and then, once the SNIa explode, both the [Mg/Mn] and [Al/Fe] decrease (thin disc). On the other hand, a system like GES that had a slower SFH \citep{fa18} starts the chemical evolution in the upper left region and as the metallicity increases, the [Al/Fe] increases slowly and remains subsolar, while the [Mg/Mn] decreases \citep{andrews17}.

We focus on the APOGEE abundances, for which the differences between the components are more clearly visible. Figure \ref{mgmn_alfe} shows the whole sample of fast rotators at all metallicities as in Figure \ref{afe}, color coded in density scale, over plotted with the fast rotators with metallicities below [Fe/H] < -1 (left panel), between -1 < [Fe/H] < -0.7 (middle panel), and -0.7 < [Fe/H] < -0.55 (right panel) in red. We see that stars with [Fe/H] < -1 clearly populate the upper left region associated to the early star formation when the stellar populations where chemically unevolved. As the metallicity increases the number of stars populating the thick disc region dominates, yet, there are stars populating the thin disc region, even at metallicities below -0.7. At higher metallicities both locations are well populated. Interestingly, at metallicities [Fe/H] > -1 there is a group of stars that follows the expected chemical evolution path of a dwarf satellite (lower-left region). These stars have also low [$\alpha$/Fe] content. Thus, they are likely candidates of being accreted stars and we excluded them to derive the MDF in Figure \ref{mdf_3}.

In summary, based on these chemical abundance distributions of the stars we observed at [Fe/H] < $-0.55$ moving in thin-disc-like orbits, we conclude that the chemo-dynamically defined thin-disc starts appearing at metallicities between -1 and -0.7. At lower metallicities the fast rotators exhibit chemical abundances that point to a star formation in a system chemically unevolved, less enriched than the bulk of the thick disc.

\subsection{A thin disc, a thick disc or a prograde halo?}

\begin{figure*}
\includegraphics[scale=0.7, trim=40 0 40 0]{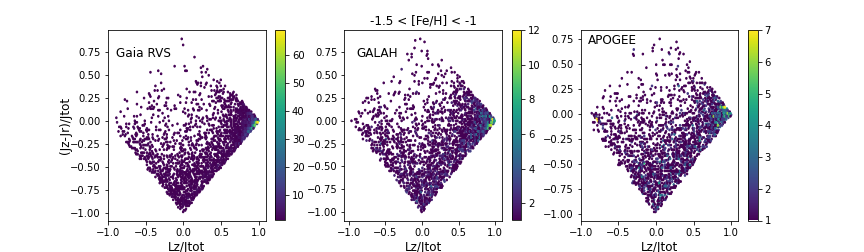}
\includegraphics[scale=0.7, trim=40 0 40 0]{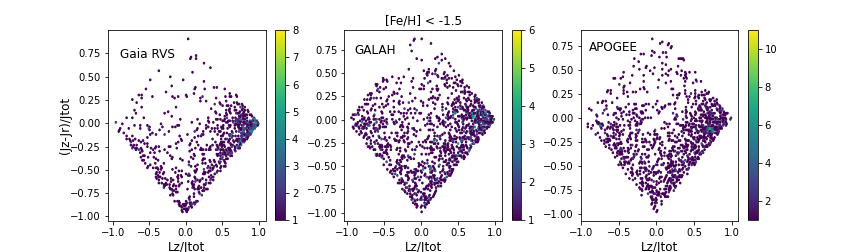}
\caption{Action space as in Figure \ref{actions} of stars with metallicities between -1.5 < [Fe/H] < -1 (top panels) and below [Fe/H] < -1.5 (bottom panels), within the Gaia RVS DR3 (left), GALAH DR3 (middle) and APOGEE DR17 (right) databases.}
\label{action_metranges}
\end{figure*}

The present analysis confirms the existence of fast rotators moving in thin-disc-like orbits below [Fe/H] < -0.55 down to [Fe/H] < -1.5 and even lower metallicities, although in fewer numbers. Their chemical trends show, however, that not all of them are part of the chemical thin disc. In particular, at metallicities [Fe/H] < -1 they present the chemical signature characteristic of an early chemical enrichment, distinct from the thick disc. In addition, their small number do not allow us to rule out that these stars could be just part of the high angular-momentum tail of another Galactic component with a broader distribution of orbits, like a prograde halo.

We would expect that, if they are part of an isotropic halo distribution, there would be the same number of stars with the same characteristics but on retrograde orbits. Figure \ref{action_metranges} shows the action space for stars within the three databases with metallicities below [Fe/H] < -1 (top panels) and [Fe/H] < -1.5 (bottom panels). There is clearly an excess of stars on prograde orbits with respect to the retrograde counterpart. The Gaia RVS plot shows a striking lack of stars compared with the other surveys, probably because the conservative selection cuts applied likely removed also \textit{bona fide} metal-poor stars from our sample. 

Recent simulations suggest that the galactic bar could induce the emergence of a stellar population on disc-like orbits initially belonging either to the halo \citep{dillamore23} or to the inner galaxy \citep{yuan23, li23}.
Our chemical analysis corroborates the fact that below [Fe/H] < -1 stars formed in a medium less chemically enriched than the bulk of the thick disc which could be the stellar halo. However, based on our results only, we cannot exclude the possibility of them being part of an ancient disc, formed before the majority of the thick disc.

The analysis performed by
\citet{zhang23}, fitting gaussian mixture models to the 3D velocity space, revealed that below [Fe/H] < -1.6 there are no signs of a significant disc population (they limit it to
$< 3 \%$ of their total distribution), which is in line with our findings. Like in this work (see the
bottom plots of Figure \ref{action_metranges}), they do find an asymmetry in rotational velocity at low metallicity (even for [Fe/H] < -2.0)
 with an excess of prograde stars, which they explained through the presence of a prograde halo component with a mean rotational velocity of
$\sim$ 80
$\rm km\ s^{-1}$. 

Further analysis with age distributions and/or more comprehensive chemical abundance measurements are required to clearly establish the precise metallicity at which the early Milky Way disc formed.

\section{Conclusions and Discussion}

In this work, we have confirmed that the metallicity distribution of stars moving in thin-disc-like orbits, i.e., with high angular momentum $\rm L_{z}/J_{tot}$ > 0.95 moving very close to the plane, within $\rm |Z_{max}|$ < 750 pc, shows a metal-poor tail that goes below the classical lower limit of [Fe/H] $\sim$ -0.7, decreasing at a constant rate in logarithmic scale down to [Fe/H] $\sim$ -1.5. Tentatively, we also found thin-disc-like stars at lower metallicities, down to [Fe/H] $\sim$ -2.5, although in much lower numbers. 

Their orbital parameters show that they are confined to the plane, with the bulk of them having $\rm |Z_{max}|<750$ pc (approximately twice the thin disc scale height, e.g. \citealt{bd16}). This orbital distribution is observed independently on the metallicity range considered or their current location in the Galaxy. In particular, they are not restricted to the anticenter region where they were found for the first time by \citetalias{FA21}, but cover the whole area surveyed by each database. However, we observed that the most metal-poor, those with [Fe/H] < -1.5, do not reach galactocentric radii larger than 10 kpc, suggesting that they do not belong to the outer disc. Dynamically selected thin disc stars with -1 < [Fe/H] < -0.5 are comprised of both high- and low-$\alpha$ populations, but stars at lower metallicities are mostly high-$\alpha$. The evaluation of the [Mg/Mn] vs. [Al/Fe] chemical space clearly shows that the chemically defined thin disc starts at metallicities between -1  < [Fe/H] < -0.7. Below [Fe/H] < -1 the fast rotators show the chemical trends typical of an earlier chemical enrichment than the bulk of the thick disc.

Using Gaia and APOGEE data, \citet{belokurov22} found that MW stars formed in-situ have a mean rotational velocity (and dispersion) variation with metallicity that could be explained by the formation of the disc from a kinematically hot spheroid of stars with a slightly net prograde rotation ($\rm V_{\phi}$  = 50 $\rm km\ s^{-1}$) that spins up to the rotation of the current disc between -1.3 < [Fe/H] < -0.9. They interpret these findings as the transition from turbulent bursty star formation through gas accretion in cold narrow filaments to a more steady gas accretion phase through cooling gas from a hot halo that allows the formation of a coherent disc. They found that these in-situ stars below [Fe/H] < $-0.9$ describe a broad velocity distribution that reaches velocities from -200 $\rm km\ s^{-1}$ to 300 $\rm km\ s^{-1}$.

Following the \citet{belokurov22} scenario our metal-poor fast rotators could be the high angular-momentum tail of a thicker disc in formation. The fact that we see an increasing number of stars as a function of the metallicity moving in thin-disc-like orbits is coherent with the possibility of an incipient disc that gains angular momentum as the metallicity increases, because we would expect the number of high angular-momentum stars to also increase \citep{kordo17}. However, based on our results, such spin up might have started at lower metallicities, at least at [Fe/H] $\sim$ -1.5.

The evaluation of other chemical species would be helpful to verify whether they could belong to an accreted population. Recently, \cite{mardini22} pointed out the existence of metal-poor disc stars with metallicities lower than -0.8 dex down to -3.5 dex, that they called the Atari disc, for which they suggested an accreted origin. This stellar group was identified by selecting stars with similar characteristics as the 'metal-weak' thick disc that had been previously detected \citep{norris85b, ruchti11, kordopatis13b, carollo19}. Although the Atari disc shares the same metallicity range with our metal-poor thin disc, there are significant differences between the two populations. The Atari disc mean rotational velocity is $<V_{\phi}>$ = 154 km $s^{-1}$, i.e., it rotates even slower than the thick disc, and, consequently, than our metal-poor thin disc. Atari stellar orbits are characterized by large eccentricities between 0.3 < $ecc$ < 0.5, typical of the halo and thick disc, contrary to our low eccentricity population, $ecc$ < 0.3. It is true that these two populations were selected with different dynamical criteria, which explains why they do not show the same dynamical properties. 

However,  the analysis of the APOGEE chemical abundances revealed that some of our selected stars may be, indeed, accreted. Further analysis will be the topic of future work.

To conclude, it is important to notice that chemical abundances derived with automatic pipelines that run on hundreds of thousands of stars should be taken with a grain of salt, since they are not thoroughly tested on peculiar cases as the one investigated in this paper. For this reason, a higher-resolution follow-up is planned to properly characterise these metal-poor thin disc stars. For this sample-to-come, reliable stellar ages will be also derived, as they are necessary to disentangle the origin of the Galactic disc populations.

\begin{acknowledgements}
      
The authors thank Patrick de Laverny for useful suggestions. EFA, GB and CG acknowledge support from the Agencia Estatal de Investigaci\'on del Ministerio de Ciencia e Innovaci\'on (AEI-MCINN) under grant "At the forefront of Galactic Archaeology: evolution of the luminous and dark matter components of the Milky Way and Local Group dwarf galaxies in the Gaia era" with reference PID2020-118778GB-I00/10.13039/501100011033. EFA also acknowledges support from the "Mar\'ia Zambrano" fellowship from the Universidad de La Laguna. GK, VH and IG gratefully acknowledge support from the french national research agency (ANR) funded project MWDisc (ANR-20-CE31-0004). AAA acknowledges support from the Herchel Smith Fellowship at the University of Cambridge and a Fitzwilliam College research fellowship supported by the Isaac Newton Trust. This research has been partially funded from a Spinoza award by NWO (SPI 78-411). This research was supported by the International Space Science Institute (ISSI) in Bern, through ISSI International Team project 540 (The Early Milky Way).\\

This work has made use of data from the European Space Agency (ESA) mission
{\it Gaia} (\url{https://www.cosmos.esa.int/gaia}), processed by the {\it Gaia}
Data Processing and Analysis Consortium (DPAC,
\url{https://www.cosmos.esa.int/web/gaia/dpac/consortium}). Funding for the DPAC
has been provided by national institutions, in particular the institutions
participating in the {\it Gaia} Multilateral Agreement.     \\

Funding for the Sloan Digital Sky 
Survey IV has been provided by the 
Alfred P. Sloan Foundation, the U.S. 
Department of Energy Office of 
Science, and the Participating 
Institutions. 

SDSS-IV acknowledges support and 
resources from the Center for High 
Performance Computing  at the 
University of Utah. The SDSS 
website is www.sdss4.org.

SDSS-IV is managed by the 
Astrophysical Research Consortium 
for the Participating Institutions 
of the SDSS Collaboration including 
the Brazilian Participation Group, 
the Carnegie Institution for Science, 
Carnegie Mellon University, Center for 
Astrophysics | Harvard \& 
Smithsonian, the Chilean Participation 
Group, the French Participation Group, 
Instituto de Astrof\'isica de 
Canarias, The Johns Hopkins 
University, Kavli Institute for the 
Physics and Mathematics of the 
Universe (IPMU) / University of 
Tokyo, the Korean Participation Group, 
Lawrence Berkeley National Laboratory, 
Leibniz Institut f\"ur Astrophysik 
Potsdam (AIP),  Max-Planck-Institut 
f\"ur Astronomie (MPIA Heidelberg), 
Max-Planck-Institut f\"ur 
Astrophysik (MPA Garching), 
Max-Planck-Institut f\"ur 
Extraterrestrische Physik (MPE), 
National Astronomical Observatories of 
China, New Mexico State University, 
New York University, University of 
Notre Dame, Observat\'ario 
Nacional / MCTI, The Ohio State 
University, Pennsylvania State 
University, Shanghai 
Astronomical Observatory, United 
Kingdom Participation Group, 
Universidad Nacional Aut\'onoma 
de M\'exico, University of Arizona, 
University of Colorado Boulder, 
University of Oxford, University of 
Portsmouth, University of Utah, 
University of Virginia, University 
of Washington, University of 
Wisconsin, Vanderbilt University, 
and Yale University.
Collaboration Overview
Affiliate Institutions
Key People in SDSS
Collaboration Council
Committee on Inclusiveness
Architects
SDSS-IV Survey Science Teams and Working Groups
Code of Conduct
SDSS-IV Publication Policy
How to Cite SDSS
External Collaborator Policy
For SDSS-IV Collaboration Members

\end{acknowledgements}

%
\bibliographystyle{aa} 
\bibliography{bibliografia.bib} 
%

\begin{appendix} 

\section{Flags quality selection on Gaia RVS.}
\label{flags}
\begin{table}[h!!]
\caption{Table of the flags and stellar parameters of the stars selected in our Gaia RVS sample.}\label{gaia_flags}
\centering
\begin{tabular}{lc}
\hline \hline
$flag\_vbroadT$ & 0,1  \\
$flag\_vbroadG$ & 0,1  \\
$flag\_vbroadM$ & 0,1  \\
$flag\_vradT$ & 0,1  \\
$flag\_vradG$ & 0,1  \\
$flag\_vradM$ & 0,1  \\
$flag\_fluxnoise$ & 0,1, and 2 where [Fe/H]\footnote{uncalibrated [M/H]}$<-2$  \\
$flag\_extrapol$ & 0,1  \\
$flag\_negflux$ & 0  \\
$flag\_nanflux$ & 0  \\
$flag\_emission$ & 0  \\
$flag\_nullfluxer$ & 0  \\
$flag\_KM$ & 0  \\
$spectraltype\_esphs$ & all except type O, B and A\\
$T_{\rm eff}$ (K) & [4250, 6000] \\
$\log g$ (dex) & [0, 4.75] \\
\hline
\end{tabular}
\footnotesize{ $^4$Uncalibrated Gaia RVS [M/H]} {\bf this is the wrong footnote number}
\end{table}

\section{Gaia RVS [Fe/H] comparison with APOGEE and GALAH [Fe/H].}
\label{ab_comp}

\begin{figure}[h!!!]
\includegraphics[scale=0.6, trim= -10 30 20 75]{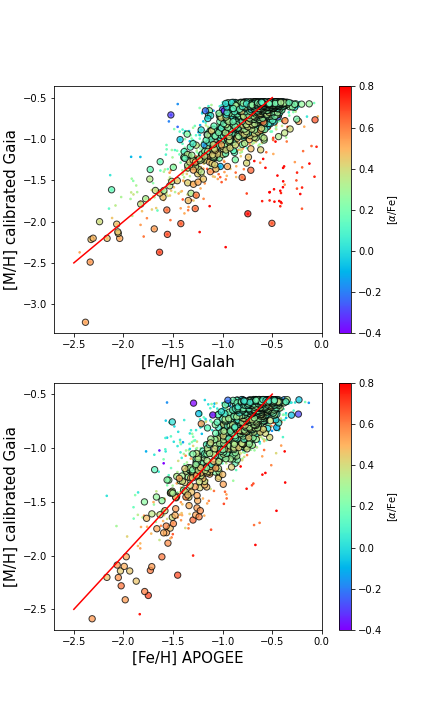}
\caption{Calibrated Gaia DR3 RVS [Fe/H] as a function of [Fe/H] estimates from GALAH DR3 (top panel) and APOGEE DR17 (bottom panel) of stars with [Fe/H] < -0.5 that pass the quality selection criteria. Small and big circles show stars before and after removing stars with Gaia DR3 RVS [Fe/H] > -2 and the $fluxnoise\_flag$ = 2, respectively. The color code indicates the [$\alpha$/Fe] Gaia RVS DR3 measurements between for each star.}
\label{ab}
\end{figure}  %

Figure \ref{ab} shows the metallicity comparison of {\bf calibrated} Gaia RVS DR3 measurements with respect to the GALAH DR3 and APOGEE DR17 determinations, color coded with the [$\alpha$/Fe] ratios measured in Gaia RVS. We see that there is a dependence of the [Fe/H] deviations with the [$\alpha$/Fe]: stars with very large [Fe/H] discrepancies correspond to objects with high [$\alpha$/Fe]. After inspection of the flags provided by the Gaia Collaboration in \citet{rb22a} we realised that the comparison of Gaia RVS [Fe/H] with APOGEE and GALAH are much improved by selecting stars with $flags\_gspspec$ chain character number 7 (hereafter $fluxnoise\_flag$) < 2. The reason why this flag helps to improve the comparison with GALAH and APOGEE estimates is because this flag warns about the flux noise induced uncertainties \citep{rb22a}. When it is equal to 2 the uncertainties in [Fe/H] are estimated to be 0.25 < $\sigma$[Fe/H] $\leq$ 0.5, being lower than 0.25 for stars flagged below < 2. 

In the case of stars with uncalibrated Gaia RVS [Fe/H] < -2 we see that, overall, after calibration, they are in fairly good agreement with the GALAH and APOGEE measurements, even those with $fluxnoise\_flag$ $=$ 2. There is, however, a negative offset in Gaia RVS [Fe/H] respect to the other surveys. In particular in GALAH, the comparison in this metallicity range shows a relatively large standard deviation. Still, because stars below [Fe/H] < -2 are scarce and that even with the negative offset and standard deviation they will remain with a metallicity below the thin disc metal-poor limit we are evaluating, [Fe/H] < -0.55, we decided to keep them all. The big circles in Figure \ref{ab} show the [Fe/H] comparison of stars after applying the final selection of flags, and the mean differences in [Fe/H] and standard deviations at several metallicity ranges are gathered in Table \ref{table_fehcomp}.

\begin{table}[h!!]
\caption{Table with the mean and standard deviations of the [Fe/H] differences between the measurements provided by Gaia RVS and those by GALAH or APOGEE, after applying all the quality cuts.}\label{table_fehcomp}
\centering
\begin{tabular}{lcc}
\hline \hline
             
Calibrated & $\rm <\Delta_{GALAH}> \pm \sigma$ & $\rm <\Delta_{APOGEE}> \pm \sigma$  \\
Gaia RVS [Fe/H] & & \\
(-3,-2) & $-0.40\pm0.63$ & $-0.30\pm0.20$ \\
(-2,-1.5) & $-0.18\pm0.33$ & $-0.15\pm0.18$ \\
(-1.5,-1) & $-0.07\pm0.20$ & $-0.01\pm0.13$ \\
(-1.5,-0.55) & $-0.02\pm0.13$ & $0.01\pm0.11$ \\

\hline
\end{tabular}
\end{table}

\section{Orbital parameters of the metal-poor thin disc stars.}
\label{app_loc}

\begin{figure*}
\includegraphics[scale=0.5,trim= 0 0 0 0]{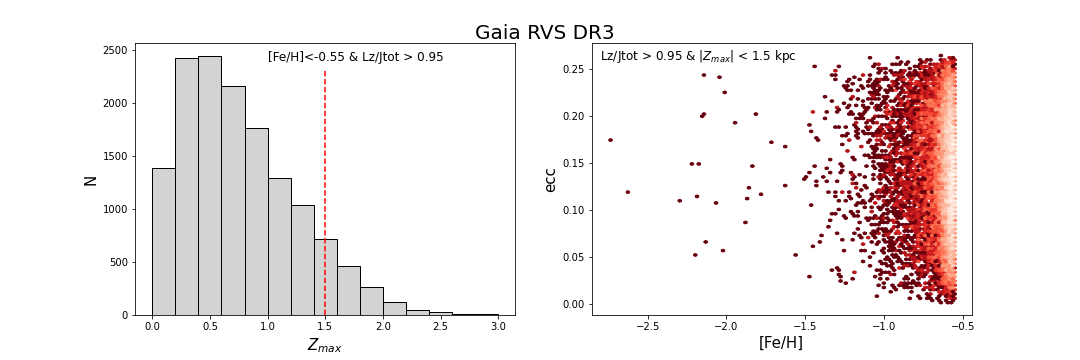}
\caption{Left panel: $\rm |Z_{max}|$ distribution of the Gaia RVS DR3 stars with [Fe/H] < -0.55 and $\rm L_{z}/J_{tot}$ > 0.95. The red dashed line at $\rm |Z_{max}| =$ 1.5 kpc correspond to the upper limit that we impose to avoid stars that reach very large distances from the Galactic plane. Right panel: Eccentricities distribution of the our final thin disc selection, i.e., stars with $\rm L_{z}/J_{tot}$ > 0.95 and $\rm |Z_{max}|$ < 1.5 kpc.}
\label{apc_fig}
\end{figure*}  %

\begin{figure*}
\includegraphics[scale=0.55,trim= 50 0 -50 0]{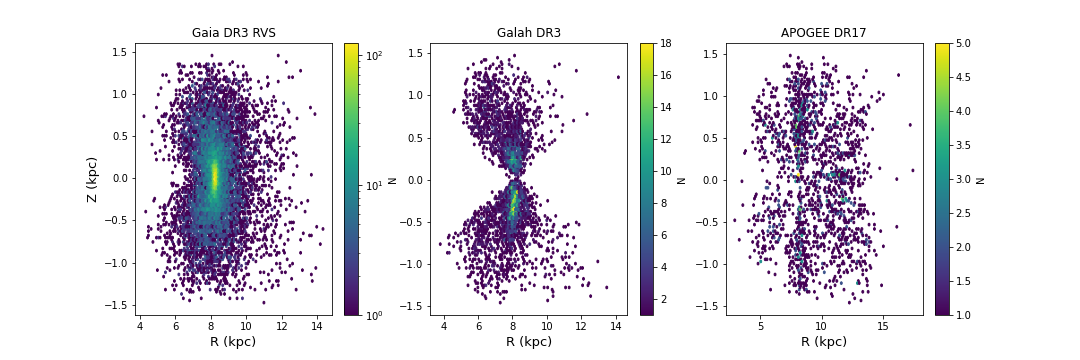}
\caption{Spatial distribution in cylindrical galactocentric coordinates of our thin disc samples selected in Gaia RVS (left), GALAH (middle) and APOGEE (right).}
\label{location}
\end{figure*}  %

The left panel of Figure \ref{apc_fig} corresponds to the distribution of $\rm |Z_{max}|$ of the stars selected in action space to have $\rm L_{z}/J_{tot}$ > 0.95. Since these selection still include stars reaching large distances from the plane we restrict our selection to $\rm |Z_{max}|$ < 1.5 kpc (red dashed line) in order to evaluate stars confined to the plane. The right panel shows the distribution of the eccentricities of our final thin disc selection of Gaia RVS DR3 stars. The eccentricity values distribute {\bf uniformly}  between 0 < $ecc$ < 0.3 and do not show trends with metallicity. 

Figure \ref{location} displays the spatial distribution of our thin disc selected stars with [Fe/H] < -0.55 gathered within Gaia DR3 RVS, GALAH DR3 and APOGEE DR17. The location of the selected thin disc stars is different depending on the survey, with Gaia DR3 RVS covering more evenly the Galactocentric radii between 6 and 10 kpc and distance from the plane between -1.5 and 1.5 kpc, thus including the anticenter explored by \citetalias{FA21}. APOGEE stars cover a similar area, but GALAH stars clearly avoid the anticenter.

\section{Impact of orbital parameters uncertainties on the metallicity distribution function.}
\label{mdf_ap_sec}

\begin{figure*}
\includegraphics[scale=0.55]{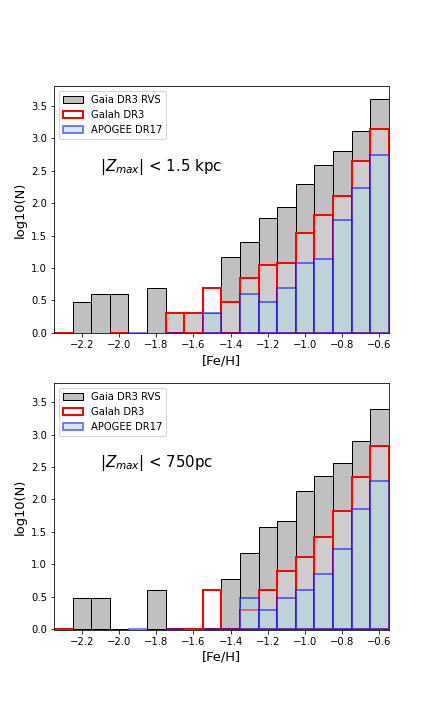}
\caption{Same as Figure \ref{mdf_3} but excluding stars for which the uncertainties on their orbital parameters leave them out of the selection criteria.}
\label{mdf_ap}
\end{figure*}  %
\end{appendix}

\end{document}